\documentclass{article}
\pdfoutput=1

\usepackage[utf8]{inputenc}
\usepackage[T1]{fontenc}
\usepackage{microtype}
\usepackage{listings}
\usepackage{graphicx}
\usepackage{balance}

\usepackage[unicode=true,hidelinks]{hyperref}

\lstdefinelanguage{Scala}%
{morekeywords={abstract,%
  case,catch,char,class,%
  def,else,extends,final,for,%
  if,import,implicit,%
  match,module,%
  new,null,%
  object,override,%
  package,private,protected,public,%
  for,public,return,super,%
  this,throw,trait,try,type,%
  val,var,%
  with,%
  yield,%
  lazy%
  },%
  sensitive,%
  morecomment=[l]//,%
  morecomment=[s]{/*}{*/},%
  morestring=[b]",%
  morestring=[b]',%
  showstringspaces=false%
}[keywords,comments,strings]%

\newcommand{\ie}{\emph{i.e.}}
\newcommand{\eg}{\emph{e.g.}}



\begin{document}

\title{Modular Remote Communication Protocol Interpreters}

\author{
  Julien Richard-Foy\\
  \texttt{julien@richard-foy.fr}
  \and
  Wojciech Pituła\\
  \texttt{w.pitula@gmail.com}
}

\maketitle

\begin{abstract}
We present \emph{endpoints}, a library that provides consistent client
implementation, server implementation and
documentation from a user-defined communication protocol description. The library
provides type safe remote invocations, and is modular and extensible. This paper
shows how its usage looks like, highlights the Scala features that its implementation
relies on, and compares it to other similar tools.
\end{abstract}

\section{Introduction}
Web applications, most smartphone applications and distributed applications have in common
that they all involve remote communication between machines. A prerequisite for remote
communication to work is that clients and servers speak the same protocol. In practice,
developers often duplicate the details of such protocols between the client implementation,
the server implementation and the public documentation (if any). This duplication
increases the risk of inconsistencies and decreases the developer
productivity.

Factoring out the parts of the protocol that are similar between clients and
servers is hard because the tasks they perform are very different, and
because they might be implemented in different projects or even different programming
languages.

We present \emph{endpoints}\footnote{Source code is available at
\url{https://github.com/julienrf/endpoints}}, a library that:

\begin{itemize}

\item turns \emph{descriptions} of
communication protocols into consistent client implementation,
server implementation and documentation;

\item guarantees that requests are well constructed: it raises a \emph{compilation error}
if one invokes an endpoint but supplies incorrect data;

\item is \emph{modular}: multiple backends are available (Play framework, scalaj-http,
akka-http, plain old XMLHttpRequest) and features can easily be made opt-in;

\item is \emph{extensible}: users can define application-specific descriptions of a particular
aspect of the underlying protocol (e.g. authorization) as well as custom interpreters
for descriptions.

\end{itemize}

In contrast with most of comparable libraries, \emph{endpoints} is written in
pure Scala and uses no macros nor code generation. Its implementation relies
on object oriented programming features such as traits and abstract type members.

The remaining of this article is organized as follows: the next section details the
problem addressed by \emph{endpoints}, section \ref{nutshell} presents the library
from a user point of view, section \ref{impl} explains how it is implemented
and section \ref{related} compares it to similar libraries.

\section{Motivation}

We can illustrate the duplication between client and server implementations of a
same protocol with the following pseudo implementation of an HTTP server
that exposes an endpoint to look up for a resource:

\begin{lstlisting}
val getItem: Request => Future[Response] = {
  case GET(p"/item/$id") =>
    itemsRepository.lookup(id)
      .map(item => Ok(item.asJson))
}
\end{lstlisting}

The corresponding pseudo implementation of a matching HTTP client could be as follows:

\begin{lstlisting}
val getItem: String => Future[Item] = id =>
  httpClient.get("/item/${urlEncode(id)}")
    .map(_.entity.fromJson[Item])
\end{lstlisting}

The details of the underlying protocol that the server and the client
have to agree on are the HTTP verb (\emph{GET}), the request path
(\emph{/item/} followed by an id), and the JSON representation of the
resource.

We observe that these elements are duplicated in the server and client
implementations. However abstracting over them is hard because
the tasks performed by the client and the server are very different.
For instance, the client uses the \texttt{id} to \emph{build} the URL
of the request, whereas the server \emph{extracts} the \texttt{id} from the URL
of an incoming request.

Furthermore, if the application has a public API, developers have to repeat
again the details of the protocol in documentation. Here is an example of an
OpenAPI\footnote{\url{https://www.openapis.org/}} definition for the \texttt{getItem} endpoint:

\begin{lstlisting}
/item/{id}:
  get:
    description: Get an item
    responses:
      200:
        description: The item identified by 'id'
        content: application/json
\end{lstlisting}

Again, we see that the HTTP verb, the URL format and the response type
are repeated in the document.

Ideally, we want users to write these elements only once and make the client,
the server and the documentation reuse them.
The next section shows how users can achieve this with the \emph{endpoints} library.

\section{\emph{endpoints} in a Nutshell}
\label{nutshell}

\subsection{Description vs Interpretation}
\label{algebra}

First, users \emph{describe} the details of the communication
protocol. For instance, the description
of our \texttt{getItem} endpoint looks as follows:

\begin{lstlisting}
val getItem: Endpoint[String, Item] =
  endpoint(
    get(path / "item" / segment[String]),
    jsonResponse[Item]
  )
\end{lstlisting}

This description defines that the used HTTP verb is \emph{GET}, the format
of the URL is \emph{/item/} followed by a text segment (the item id),
and that the response uses the JSON content-type.

The type of the \texttt{getItem} member is \texttt{Endpoint[String, Item]}. As
this will be illustrated in the next section, this type is abstract and has a
different meaning in the context of the client, the server or the
documentation.

\subsection{Type Safe Remote Calls}

The library provides a client implementation for the above endpoint
description. Its usage looks as follows:

\begin{lstlisting}
val item: Future[Item] = getItem("abc123")
\end{lstlisting}

From the client point of view, \texttt{getItem} is a \emph{function}
that takes as parameter the required information to build the request (in
our case, the item identifier) and eventually returns the \texttt{Item} instance.

The \texttt{getItem} implementation encodes the parameters (the item id), performs the HTTP request
and decodes the JSON response. If a user tries to invoke the \texttt{getItem} function
with a parameter of
the wrong type (e.g. an \texttt{Int}), a compilation error is raised.

The \emph{endpoints} library also provides a server for the endpoint description.
Its usage looks as follows:

\begin{lstlisting}
getItem.implementedByAsync { (id: String) =>
  itemsRepository.lookup(id)
}
\end{lstlisting}

From the server point of view, \texttt{getItem} is an object that has an\\
\texttt{implementedByAsync} method, which takes as parameter the business logic
performing the lookup. The result of the \texttt{implementedByAsync} call is a
\emph{request handler}: a function that decodes information from incoming
requests that match the URL and HTTP verb of the endpoint description, invokes the
business logic, and
builds an HTTP response by encoding the returned \texttt{Item} instance in JSON.

We see that with the \emph{endpoints} library, the details of the underlying
communication protocol are not duplicated anymore between the client and the
server.

\subsection{Modularity}

In the above example, the \texttt{getItem} definition uses methods
(\texttt{endpoint}, \texttt{path}, \texttt{jsonResponse}, etc.) that are
provided by traits. The whole definition looks as follows:

\begin{lstlisting}
trait Descriptions
    extends Endpoints with JsonEntities {
  val getItem = ...
}
\end{lstlisting}

The traits \texttt{Endpoints} and \texttt{JsonEntities} provide \emph{vocabulary}
(methods) to describe endpoints.

The client implementation extends the \texttt{Descriptions} trait and attaches a
particular \emph{meaning} to the vocabulary used to describe the endpoints:

\begin{lstlisting}
object Client extends Descriptions
    with xhr.Endpoints with xhr.JsonEntities
\end{lstlisting}

The \texttt{xhr} package provides traits that implement the
\texttt{Endpoints} and\\
\texttt{JsonEntities} traits with a semantics of a client. They use
the \texttt{XMLHttpRequest} API\footnote{The XMLHttpRequest API is available in web browsers and
is used to perform HTTP requests.} for communicating with the server.

The following diagram illustrates this architecture.

\begin{center}
\includegraphics[width=8cm]{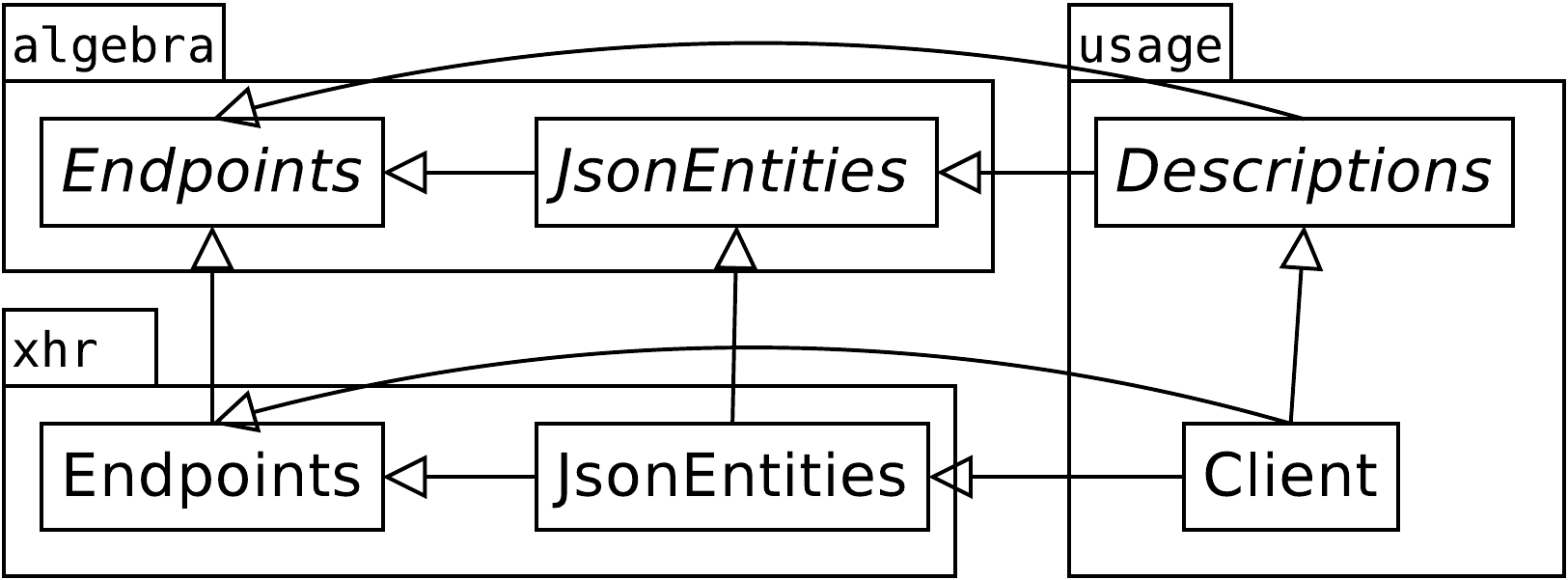}
\end{center}

The fact that the client interpreter (in the \texttt{xhr} package) is decoupled
from the vocabulary (in the \texttt{algebra} package) makes it easy, at use site,
to isolate the endpoint descriptions in a module and then have the client,
server or other interpreters in separate modules. Also, each interpreter can
bring its own dependencies without affecting the descriptions.

In our example the \texttt{Endpoints} trait provides the base vocabulary to
describe endpoints, and the \texttt{JsonEntities} enriches this vocabulary
with concepts related to the JSON format. The \emph{endpoints}
library provides several other traits, each one providing vocabulary for
a specific concern (e.g. authentication). We call such traits \emph{language
units}. The more language units a user uses, the more features are available,
but at the same time, the higher the effort to implement an interpreter
for these features. The fact that language units are in separate traits
allows us to implement \emph{partial interpreters}: interpreters for
a subset of the available language units.

\subsection{Extensibility}

Thanks to this modular design, users can easily introduce their own
application-specific language units. To achieve this they first define
a trait with abstract members introducing new vocabulary specific to
their application:

\begin{lstlisting}
trait Authorization extends Endpoints {
  type Authorized[A] = Response[Option[A]]
  def authorized[A](response: Response[A]): Authorized[A]
}
\end{lstlisting}

This trait might extend an existing language unit, depending
on whether the user wants to reuse or not the features provided by this language
unit.

Then, users can implement various interpreters for their language unit,
by defining a trait that implements its abstract members:

\begin{lstlisting}
trait ClientAuthorization
  extends Authorization with xhr.Endpoints ...
\end{lstlisting}

Again, an interpreter might reuse an existing interpreter
infrastructure. In the above example, we reuse the \texttt{xhr} interpreter,
so that our \texttt{ClientAuthorization} interpreter can be
mixed to an existing stack of \texttt{xhr} interpreters to add the
ability to interprete the \texttt{Authorization} language unit.

\subsection{First Class Citizen Descriptions}

In contrast with some similar tools (see section \ref{related}), language units are
embedded DSLs\cite{hudak1996building}: descriptions
of endpoints are not purely data (e.g. a JSON or XML dialect), but they are
Scala code, which means that all the means of abstraction of the Scala language
can be freely used. For instance, one can define a \texttt{val} to hold
a part of a description that is going to be reused for several endpoint definitions.

\section{Implementation}
\label{impl}
\subsection{Overall Pattern}

In essence, language units and interpreters are algebra interfaces and object algebras
\cite{Oliveira2012}, respectively. Concepts introduced by language units (\eg{} request,
URL, query string) are modeled with a corresponding abstract type. In practice the
\texttt{Endpoints} language unit defines 12 such concepts. In order to simplify
the signature of traits defining language units, we use the encoding proposed by
Hofer \emph{et. al.}\cite{Hofer2008}
where type carriers are modeled with abstract type members instead
of type parameters.

For instance, the \texttt{Endpoints} trait that introduces an abstract method \texttt{get},
which returns a \texttt{Request[A]}, is defined as follows:

\begin{lstlisting}
package algebra
trait Endpoints {
  type Url[A]
  type Request[A]
  def get[A](url: Url[A]): Request[A]
}
\end{lstlisting}

We omitted the methods that make it possible to create \texttt{Url[\_]} instances for
the sake of brevity.

The type \texttt{Request[A]} represents a request that carries an information of type
\texttt{A}. For instance, a request containing an id of type \texttt{Long} would
have type \texttt{Request[Long]}. Similarly, a request containing two parameters,
one of type \texttt{Int} and one of type \texttt{String}, would have type
\texttt{Request[(Int, String)]}, and so on for other arities.

The method \texttt{get} describes an HTTP request that uses the verb \emph{GET}.

In summary, language units introduce \emph{concepts} (abstract type members) and ways
to \emph{create} or \emph{combine} them (abstract methods).

Interpreters give a concrete meaning to these concepts. For instance, the \texttt{xhr}
interpreter embodies the client side of a request:

\begin{lstlisting}
package xhr
trait Endpoints extends algebra.Endpoints {
  type Request[A] = A => XMLHttpRequest
}
\end{lstlisting}

We omitted the implementation of the methods for the sake of brevity.

From the point of view of this client, a \texttt{Request[A]} is a function
that takes an \texttt{A} and returns an \texttt{XMLHttpRequest} instance. In this
case, the \texttt{A} information is what is needed to build the request.

The server interpreter is defined like so:

\begin{lstlisting}
package playserver
trait Endpoints extends algebra.Endpoints {
  type Request[A] = RequestHeader => Option[BodyParser[A]]
}
\end{lstlisting}

For this server interpreter (backed by Play framework), a \texttt{Request[A]}
is a function checking that an incoming request matches the description, and in such
a case returns an object that extracts an \texttt{A} from the request.

\subsection{Forwarding Interpreters}

As mentioned in the introduction, the \emph{endpoints} library is also able to
generate an OpenAPI documentation from endpoint descriptions. However, the language units
used to describe the endpoints can not be the same as those used so far, because
documentation needs additional information (such as human-readable
descriptions).

As an example, here is how one would write the \texttt{getItem} definition using these
language units:

\begin{lstlisting}
val getItem: Endpoint[String, Item] =
  endpoint(
    get(path / "item" / segment[String]("id")),
    jsonResponse[Item](
      documentation = "The item identified by 'id'"
    )
  )
\end{lstlisting}

This listing is similar to the one given in section \ref{algebra} but some methods
take additional parameters (e.g. there is a
\texttt{documentation} parameter in the \texttt{jsonResponse} method) that
are used to generate the documentation.

Since the language unit is different from the one presented in section \ref{algebra},
we can not apply
the client and server interpreters to it. However, being able to reuse these
interpreters on such descriptions is useful, that’s why we implemented
a solution based on forwarding\cite{buchi2000generic}:

\begin{center}
\includegraphics[width=8cm]{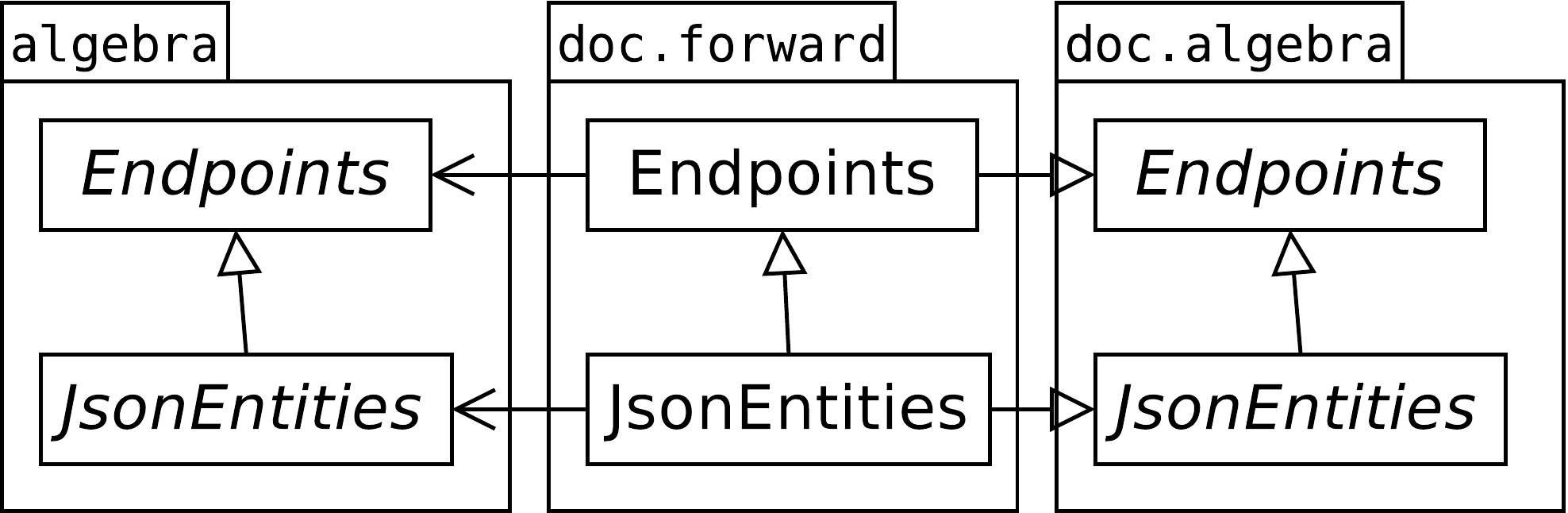}
\end{center}

The \texttt{doc.forward} package provides interpreters for
\texttt{doc.algebra} language units. These interpreters are implemented by forwarding calls to an
\texttt{algebra} interpreter. It is worth noting that this relationship
can be refined when language units are refined (\ie{} the \texttt{receiver} member of the
\texttt{doc.forward.JsonEntities} interpreter refers to an \texttt{algebra.JsonEntities}
interpreter).

For instance, here is an excerpt of the implementation of the\\
\texttt{doc.forward.Endpoints} interpreter:

\begin{lstlisting}
trait Endpoints extends doc.algebra.Endpoints {
  val receiver: algebra.Endpoints
  type Request[A] = receiver.Request[A]
  type Url[A] = receiver.Url[A]
  def get[A](url: Url[A]): Request[A] = receiver.get(url)
}
\end{lstlisting}

For users, applying the \texttt{xhr} client intepreter to documented
descriptions looks as follows:

\begin{lstlisting}
object Client extends DocDescriptions
    with forward.Endpoints
    with forward.JsonEntities {
  lazy val receiver = new xhr.Endpoints with xhr.JsonEntities
}
\end{lstlisting}

\section{Related Works}
\label{related}
\subsection{Autowire / Remotely / Lagom}

Autowire\footnote{\url{https://github.com/lihaoyi/autowire}} and
Remotely\footnote{\url{http://verizon.github.io/remotely}} are Scala libraries automating
remote proceduce calls between a server and a client.
Lagom\footnote{\url{https://www.lagomframework.com/}} is
a framework for implementing microservices. One difference with \emph{endpoints} is
that these tools are based on
macros generating the client according to the interface (defined as a Scala trait)
of the server. These macros make these solutions harder to reason about (since
they synthesize code that is not seen by users) and their implementation might not
support all edge cases\footnote{Several issues have been reported about macro expansion:
\url{https://goo.gl/Spco7u}, \url{https://goo.gl/F2E5Ev} and \url{https://goo.gl/LCmVr8}}.

A more fundamental difference is that in Autowire and Remotely, the underlying HTTP communication
is viewed as an implementation detail, and all remote calls are multiplexed through
a single HTTP endpoint. In contrast, the goal of \emph{endpoints} is to embrace the features
of the HTTP protocol (content negotiation, authorization, semantic verbs and status codes,
etc.), so, in general, one HTTP endpoint is used for one remote call (though the library also
supports multiplexing in case users don’t care about the underlying HTTP protocol).

Last, Autowire, Remotely and Lagom can not generate documentation of the commmunication protocol.

\subsection{Swagger / Thrift / Protobuf}

Solutions such as Swagger, Thrift and Google Protocol Buffers generate
the client and server code based on a protocol definition.
We believe that generated code is hard to reason about and to integrate and keep
in sync with code written by developers.
Also, the protocol is defined in a dedicated language (JSON dialect or custom language) which
is not extensible and not as convenient as using a fully-blown programming language like Scala.

\subsection{Rho / Fintrospect}

Fintrospect\footnote{\url{http://fintrospect.io/}} and
Rho\footnote{\url{https://github.com/http4s/rho}} are the libraries closest to \emph{endpoints}.
Their features and usage are similar: users describe their communication protocol in plain
Scala and the library generates client (Fintrospect only), server and documentation.
The key difference is that the communication protocol is described by a sealed AST,
which is not extensible: users can not extend descriptions with application-specific concerns
and interpreters can not be partial.

\subsection{Servant}

Servant\cite{mestanogullari2015type} is a Haskell library that uses generic programming to
derive client, server and documentation from endpoint descriptions. The descriptions and
interpreters are extensible. The difference with \emph{endpoints} is that in
Servant descriptions are \emph{types}, whereas in \emph{endpoints} they are \emph{values}.

Using types as descriptions has some benefits: they can directly be used to type instances of
data (in contrast, in \emph{endpoints} descriptions of data types have to mirror a
corresponding type definition). On the other hand, we believe that abstracting and combining
types using type-level computations is, in general, less convenient for users.

\section{Conclusion}

We presented \emph{endpoints}, a modular and extensible library to
perform remote communication in Scala. Its implementation mainly relies on
the object oriented features of the language such as abstract methods,
abstract type members and multiple inheritance.

\balance

\bibliographystyle{plain}
\bibliography{endpoints}

%

\end{document}